\documentclass[twocolumn,superscriptaddress,amsmath,ammsymb]{revtex4-2}
\usepackage{epsfig}
\usepackage{amsmath}
\usepackage{amssymb}
\usepackage{graphicx}
\usepackage{dcolumn}
\usepackage{bm}
\usepackage[usenames]{color}
\usepackage[breaklinks]{hyperref}
\hypersetup{colorlinks=true,allcolors=blue}

\begin{document}
	
\title{Gapped Dirac semimetal with mixed linear and parabolic dispersions}

\author{Yi-Xiang Wang}
\email{wangyixiang@jiangnan.edu.cn}
\affiliation{School of Science, Jiangnan University, Wuxi 214122, China}
\affiliation{School of Physics and Electronics, Hunan University, Changsha 410082, China}

\author{Fuxiang Li}
\email{fuxiangli@hnu.edu.cn}
\affiliation{School of Physics and Electronics, Hunan University, Changsha 410082, China}

\date{\today}

\begin{abstract}

In this paper, we make a comprehensive study of the properties of a gapped Dirac semimetal model, which was originally proposed in the magnetoinfrared spectroscopy measurement of ZeTe$_5$, and includes both the linear and parabolic dispersions in all three directions.  We find that, depending on the band inversion parameters, $\zeta'$ and $\zeta_z'$, the model can support three different phases: the single Dirac point (DP) phase, the double DPs phase and the Dirac ring phase.  The three different phases can be distinguished by their low-energy features in the density of states (DOS) and optical conductivity.  At high energy, both the DOS and optical conductivity exhibit power-law like behaviors, with the asymptotic exponents depending heavily on the signs of $\zeta'$ and $\zeta_z'$.  Moreover, the thumb-of-rule formula between the DOS and optical conductivity is satisfied only when $(\zeta',\zeta_z')>0$.  The implications of our results for experiments are discussed. 
\end{abstract} 

\maketitle

\section{Introduction} 
  
The topological phases of matter possess topological properties that are insensitive to the external perturbations and are protected by symmetries.  Among the search of the topological materials in the past few years, including topological insulators (TIs) ~\cite{M.Z.Hasan, X.L.Qi, A.Bansil}, three-dimensional (3D) Dirac and Weyl semimetals~\cite{N.P.Armitage}, the single crystal ZrTe$_5$ has excited ongoing interests as a candidate for the novel topological material.  The early \textit{ab initio} calculations indicated that the single-layer ZrTe$_5$ was a good two-dimensional (2D) TI with a large energy gap of 100 meV~\cite{H.Weng}.  It was also predicted that the bulk ZrTe$_5$ was revealed to be near the phase boundary between the weak and strong TIs~\cite{H.Weng}.  However, in experiment, the ground state of bulk ZrTe$_5$ is still under debate.  Different studies suggested that it might be a 3D Dirac semimetal~\cite{Q.Li, G.Zheng, R.Y.Chen2015a, R.Y.Chen2015b, B.Xu, Z.G.Chen, Y.Liu}, a weak TI~\cite{L.Moreschini, X.B.Li, R.Wu, Y.Zhang} or even a strong TI~\cite{G.Manzoni}.  The complication may be attributed to the fact that the system is close to the band inversion point~\cite{H.Weng} and is also sensitive to the purity of the crystal, temperature~\cite{B.Xu} and pressure~\cite{H.Weng, Y.Zhang, J.L.Zhang, Y.Zhou}. 
 
In the early experiments, it was commonly accepted that the low-energy excitations in ZrTe$_5$ could be captured by a 3D massive/massless Dirac model~\cite{R.Y.Chen2015b,Q.Li,Z.G.Chen}, as evidenced by the linear dependence of the optical conductivity on the photon frequency over a range of the wave number 50$\sim$1200 cm$^{-1}$~\cite{R.Y.Chen2015a,B.Xu}.  However, the recent experiments questioned the dimensionality of the Dirac cone in ZrTe$_5$.  An optical spectroscopy study pointed out that the Dirac cone was only 2D and put forward a 2D conical model to describe its ground state~\cite{E.Martino}.  The following theoretical studies revealed that the optical conductivities in the 2D conical model exhibit different dependence on the photon frequency along the $x-$ and $z-$direction, as Re$(\sigma_x)\propto\omega^\frac{1}{2}$ and Re$(\sigma_z)\propto\omega^\frac{3}{2}$~\cite{Y.X.Wang2020, Z.Rukelj}.  Another magnetoinfrared spectroscopy measurement~\cite{Y.Jiang} by Jiang \textit{et al.}
suggested that its ground state was a strong topological insulator and could be described by a gapped Dirac semimetal model, in which both the linear and parabolic dispersions are mixed in all three directions, with the most pronounced evidence was the observation of an additional set of optical transitions from a band gap next to the Brillouin center.  The nonlinear low-energy dispersions in ZrTe$_5$ are also supported by the angle-resolved photoemission spectroscopy results~\cite{T.Liang} as well as the Shubnikov-de Haas effect\cite{Y.Liu,Z.Sun}. 

Motivated by these progresses, in this paper, we make a detailed study on the proposed gapped Dirac semimetal model~\cite{Y.Jiang} as well as on how to characterize the bulk electronic states.  We calculate the density of states (DOS) and the interband optical conductivity, with the influence of the impurity scatterings being included phenomenologically.  Our main results are as follows: (i) Depending on the band inversion parameters, $\zeta'$ and $\zeta_z'$, we find that the gapped Dirac semimetal model can support three different phases: the single Dirac point (DP) phase, the double DPs phase and the Dirac ring phase, as shown in the phase diagram of  Fig.~\ref{Fig1}.  (ii) The different phases can be distinguished by their low-energy features in the DOS and optical conductivity.  (iii) At high energy, both the DOS and optical conductivity exhibit the power-law like behaviors, which can be captured by asymptotic exponents.  We extract the asymptotic exponents by fitting the numerical data and find that they depend heavily on the signs of $\zeta'$ and $\zeta_z'$.  More interestingly, the rule-of-thumb formula between the DOS and optical conductivity is only satisfied when $(\zeta',\zeta_z')>0$.  (iv) At high frequency, the optical conductivity can exhibit certain robustness to the impurity scatterings when $\zeta'\zeta_z'>0$.  Our work can help understand the ground state properties in bulk ZrTe$_5$ and may also shed lights on the future study of the topological electronics.

\section{Model and phase diagram}

\begin{figure}
	\includegraphics[width=8.2cm]{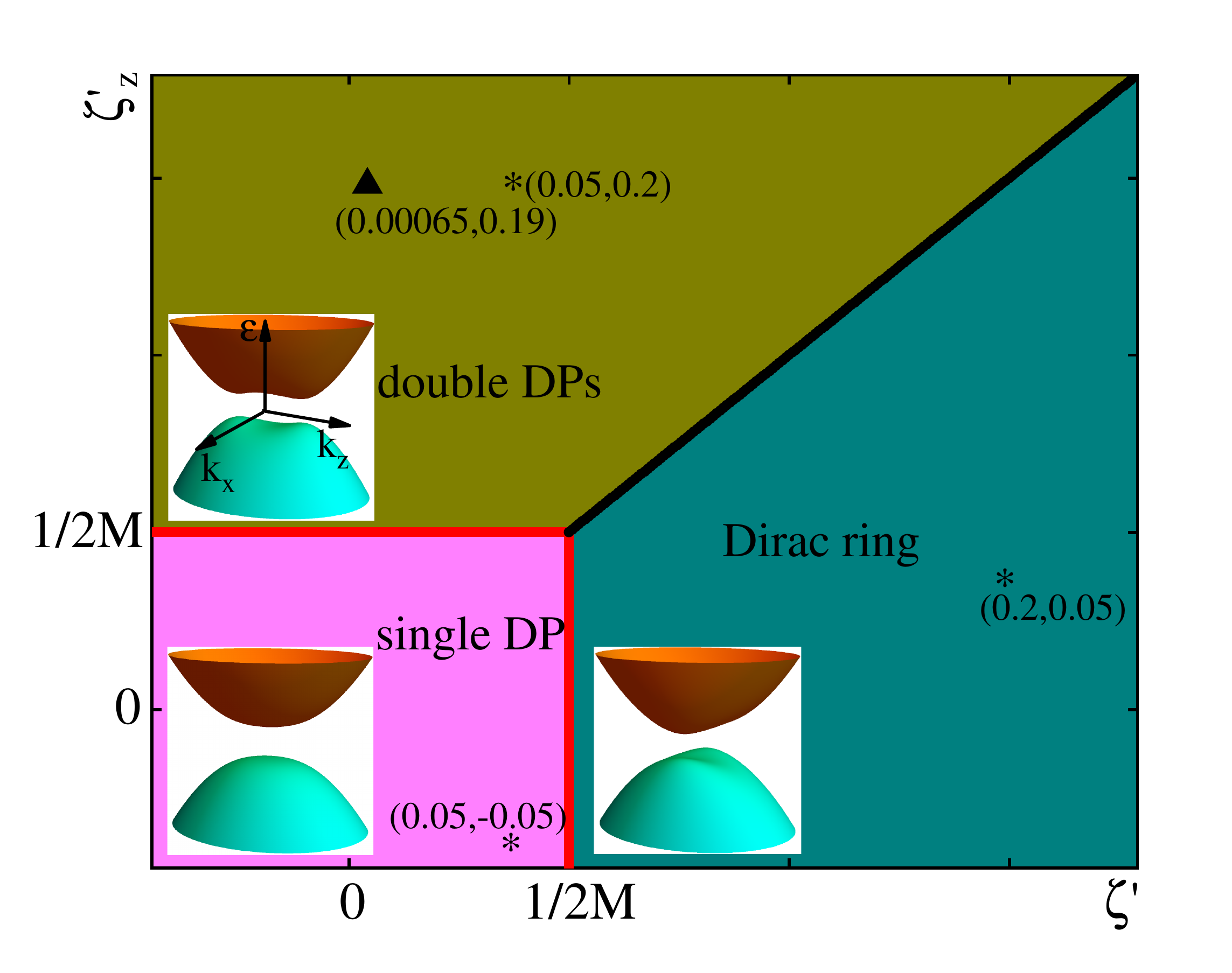}
	\caption{(Color online) Phase diagram described by $H(\boldsymbol k)$ in the parameter space $(\zeta',\zeta_z')$.  Three phases exist and are denoted in different colors, including the single DP phase, the double DPs phase and the Dirac ring phase.  The schematics of the energy bands are given in the $(k_x,k_z)$ space with $k_y=0$.  The red lines denote the phase boundaries at which the Dirac points are reduced to lie at the $\Gamma$ point.  The asterisks label the parameters used in the following calculations, and the triangle labels the parameters extracted from the experiment~\cite{Y.Jiang}.  The Dirac mass is taken as $M=7.5$ meV. }
	\label{Fig1}
\end{figure}

We start from the model that was proposed to explain the magnetoinfrared spectroscopy in ZrTe$_5$~\cite{Y.Jiang}.  In the four-component basis of $(|+,\uparrow\rangle, |+,\downarrow\rangle, |-,\uparrow\rangle, |-,\downarrow\rangle)^T$, the Hamiltonian is written as ($\hbar=1$), 
\begin{align}
H(\boldsymbol k)=&v k_x\tau_x\otimes\sigma_z
+v k_y\tau_y\otimes I  
+v_z k_z\tau_x\otimes\sigma_x 
\nonumber\\
&+L\tau_z\otimes I.  
\end{align}  
Here $L=M-\zeta(k_x^2+k_y^2)-\zeta_z k_z^2$.  $v$ and $v_z$ are the Fermi velocities, $M$ denotes the Dirac mass, $\zeta$ and $\zeta_z$ are the band inversion parameters.  $\tau$ and $\sigma$ are the Pauli matrices acting on the orbital and spin degree of freedom, respectively.  For simplicity, we have assumed that $v_x=v_y=v$ and $\zeta_x=\zeta_y=\zeta$.  $H(\boldsymbol k)$ preserves the time-reversal symmetry ${\cal T}^{-1}H(-\boldsymbol k){\cal T}=H(\boldsymbol k)$, with ${\cal T}=i\sigma_y {\cal K}$ and $\cal K$ being the complex conjugation operator, and the inversion symmetry ${\cal I}^{-1} H(-\boldsymbol k) {\cal I}=H(\boldsymbol k)$, with ${\cal I}=\tau_z$.  The particle-hole symmetry is preserved only when the first (second) term is absent, ${\cal C}^{-1} H(-\boldsymbol k){\cal C}=-H(\boldsymbol k)$, with ${\cal C}=\tau_y(\tau_x)$.  So according to the Altland and Zirnbauer notations, the system described by $H(\boldsymbol k)$ belongs to the symplectic class AII~\cite{A.P.Schnyder, C.K.Chiu}. 

By direct diagonalizing $H(\boldsymbol k)$, the eigenenergies are given as
\begin{align}
\varepsilon_{1,2}(\boldsymbol k)=-\varepsilon_{3,4}(\boldsymbol k) =\varepsilon(\boldsymbol k)=\sqrt{v^2 k_\parallel^2+v_z^2 k_z^2+L^2},  
\end{align} 
which are twofold degenerate, with the in-plane wave vector $k_\parallel=\sqrt{k_x^2+k_y^2}$.  The corresponding eigenvectors are
\begin{align}
&\psi_1=c_+\begin{pmatrix}
\frac{v_z k_z}{\varepsilon+L}\\ 
\frac{-vk_+}{\varepsilon+L}\\
0\\-1
\end{pmatrix}, \quad
\psi_2=c_+\begin{pmatrix}
\frac{vk_-}{\varepsilon+L}\\ 
\frac{v_z k_z}{\varepsilon+L}\\
-1\\ 0
\end{pmatrix},
\nonumber\\ 
&\psi_3=c_-\begin{pmatrix}
\frac{v_z k_z}{\varepsilon-L}\\ 
\frac{-vk_+}{\varepsilon-L}\\
0\\ 1
\end{pmatrix}, \quad
\psi_4=c_-\begin{pmatrix}
\frac{vk_-}{\varepsilon-L}\\ 
\frac{v_z k_z}{\varepsilon-L}\\ 
1\\ 0
\end{pmatrix}, 
\end{align}
with the normalized coefficients $c_\pm=\sqrt{\frac{\varepsilon\pm L}{2\varepsilon}}$ and $k_\pm=k_x\pm ik_y$.  The Dirac points (DPs) of $H(\boldsymbol k)$ are determined by the minima of $\varepsilon$.  We find that depending on the model parameters, $H(\boldsymbol k)$ can support three kinds of phases that are classified by the DPs~\cite{Supp}: (i) the single DP phase, with the DP $(k_\parallel,k_z)=(0,0)$ at the center of the Brillouin and the gap $\Delta_s=2M$, (ii) the double DPs phase, with the DPs $(k_\parallel,k_z)=(0,\pm\sqrt{\frac{M}{\zeta_z}-\frac{v_z^2}{2\zeta_z^2}})$ in the $z-$direction and the gap $\Delta_d=2\sqrt{\frac{v_z^2M}{\zeta_z}-\frac{v_z^4}{4\zeta_z^2}}$, (iii) the Dirac ring phase with the DPs $(k_\parallel,k_z)=(\sqrt{\frac{M}{\zeta}-\frac{v_z^2}{2\zeta^2}},0)$, which form a circular line in the $k_x-k_y$ plane, and the gap $\Delta_r=2\sqrt{\frac{v^2M}{\zeta}-\frac{v^4}{4\zeta^2}}$. 

In Fig.~\ref{Fig1}, the phase diagram is plotted in the parameter space $(\zeta',\zeta_z')$, where the rescaled quantities are $\zeta'=\frac{\zeta}{v^2}$ and $\zeta_z'=\frac{\zeta_z}{v_z^2}$.  In the phase diagram, the single DP phase dominates when max$(\zeta',\zeta_z')<\frac{1}{2M}$, the double DPs phase dominates when $\zeta_z'>$max$(\zeta',\frac{1}{2M})$, and the Dirac ring phase dominates when $\zeta'>$max$(\zeta_z',\frac{1}{2M})$.  The three phases meet at the tricritical point $(\zeta',\zeta_z')=(\frac{1}{2M},\frac{1}{2M})$.  At the phase boundaries labeled by the red lines, the DPs are reduced to lie at the $\Gamma$ point.  In the insets of Fig.~\ref{Fig1}, the energy bands of the three phases are schematically plotted in the $(k_x,k_z)$ space with $k_y=0$.  It shows that the $\Gamma$ point acts as an energy minimum in the single DP phase, but as a local energy maximum in another two phases, leading to different characteristic low-energy DOS and optical conductivity, as discussed below.  

\begin{figure*}
	\centering
	\includegraphics[width=18.2cm]{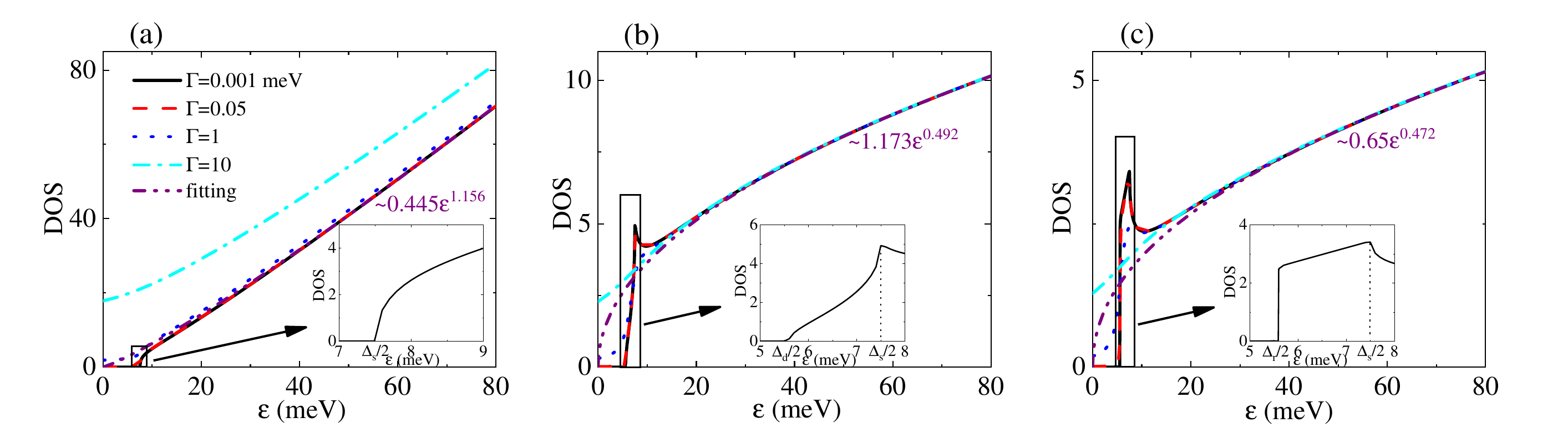}
	\caption{(Color online) The DOS (in unit of $\frac{\text{meV}^2}{v^2v_z}$) in different phases, with the chosen parameters being labeled by the asterisks in Fig.~\ref{Fig1}. The insets show the enlarged section of the low-energy DOS with the very low linewidth $\Gamma=0.001$ meV, which actually corresponds to the limiting clean case. We fit the DOS at high energy $\varepsilon\gg M$ with the power function $D(\varepsilon)=a_1\varepsilon^b$ to extract the exponent $b$.  The different $\Gamma$ are considered in the DOS and the legends are the same. }
	\label{Fig2}
\end{figure*}

In Ref.~\cite{Y.Jiang}, the parameters extracted from the magnetoinfrared spectroscopy experiment are $M=7.5$ meV, $(v,v_z)\simeq(6,0.5)\times10^5$ m/s, $(\zeta,\zeta_z)\simeq(0.1,0.2)$ eV$\cdot$nm$^2$.  The rescaled parameters $(\zeta',\zeta_z')\simeq(6.5\times10^{-4},0.19)$ meV$^{-1}$ and are denoted by the triangle in the phase diagram, pointing to the double DPs phase of the ZrTe$_5$ sample in their experiment.  

It is worth noting that there are actually several well-established models that are the limiting cases of our model Hamiltonian $H(\boldsymbol k)$.  First, when $M=0$ and $(\zeta',\zeta_z')=(0,0)$, $H(\boldsymbol k)$ reduces to the simplest Dirac semimetal model, which can describe the single DP phase in Na$_3$Bi~\cite{Z.Liua} and  Cd$_3$As$_2$~\cite{S.Borisenko, Z.Liub}. The second is the 2D conical model of ZrTe$_5$~\cite{E.Martino, Y.X.Wang2020, Z.Rukelj}, where only the linear term in the $x-y$ plane and the parabolic term in the $z-$direction are included.  So the model is equivalent to $H(\boldsymbol k)$ when $\zeta=v_z=0$ and corresponds to the point of $(\zeta',\zeta_z')=(0,\infty)$ in the phase diagram.  In addition, for the Dirac nodal line semimetal model~\cite{Y.H.Chan, W.B.Rui} that was used to describe the low-energy excitations in Ca$_3$P$_2$, the linear term in the $x-y$ plane is absent when compared to $H(\boldsymbol k)$.  The estimated parameters in Ca$_3$P$_2$~\cite{Y.H.Chan, T.Matsushita} are $M=0.37$ meV and $(\zeta',\zeta_z')=(\infty,6.9\times10^{-4})$ meV$^{-1}$.  Therefore, all the three phases in our phase diagram have their experimental correspondences.

\section{DOS and optical conductivity}

We can calculate the DOS of the system with the help of the Green's function $G(\boldsymbol k,z)=[z-H(\boldsymbol k)]^{-1}$, which is given as 
\begin{align}
D(\varepsilon)=-\frac{1}{\pi V}\sum_{\boldsymbol k}\text{Im} G(\boldsymbol k,z=\varepsilon+i\Gamma). 
\end{align}
Here $V$ is the volume of the system, $\Gamma=\tau^{-1}$ represents the linewidth broadening and $\tau$ is the scattering rate.  We use $\frac{\text{meV}^2}{v^2v_z}$ as the unit of the DOS, which can be seen in Eqs.~(10) and (11) below.  Note that $\Gamma$ is assumed as a constant in the calculations. 

We also consider the interband optical conductivity, which is obtained by using the linear-response Kubo's formula~\cite{Y.X.Wang2020}, 
\begin{align}
\text{Re}(\sigma_\alpha) 
&=-\frac{1}{V}\sum_{\boldsymbol k m\neq n}
\frac{f(\varepsilon_{m\boldsymbol k})-f(\varepsilon_{n\boldsymbol k})}
{\varepsilon_{m\boldsymbol k}-\varepsilon_{n\boldsymbol k}}
|\langle\psi_{m\boldsymbol k}|J_{\boldsymbol k\alpha}|\psi_{n\boldsymbol k}\rangle|^2
\nonumber\\
&\times\delta_\Gamma(\omega+\varepsilon_{m\boldsymbol k}-\varepsilon_{n\boldsymbol k}). 
\end{align}
Here $\alpha=x,z$ is the direction that the optical field acts on, $\omega$ is the photon frequency, $f(\varepsilon)$ is the Fermi-Dirac distribution function, $J_{\boldsymbol k \alpha}$ is the $\alpha-$direction current density operator with $r_\alpha$ denoting the position operator, and $\delta_\Gamma$ is the Lorentzian function with the linewidth $\Gamma$.  In the following, we consider the zero temperature and zero Fermi energy, $T=\varepsilon_F=0$. 

To calculate the optical conductivity, it is more convenient to change from the Cartesian into the Cylindrical coordinates, with the substitutions $\rho\text{cos}\theta=vk_x$, $\rho\text{sin}\theta=vk_y$, and $z=v_zk_z$.  For the matrix elements of $J_{\boldsymbol k\alpha}$, we have $|J_{\boldsymbol k\alpha}^{13}|=|J_{\boldsymbol k\alpha}^{24}|\neq|J_{\boldsymbol k\alpha}^{14}|=|J_{\boldsymbol k\alpha}^{23}|$, meaning that the matrix elements are not exactly equal for the degenerate eigenstates.  After completing the integration with $\theta$, we obtain Re$(\sigma_\alpha)$=2Re$(\sigma_\alpha^{13})$+2Re$(\sigma_\alpha^{14})$, in which Re$(\sigma_\alpha^{mn})$ denotes the $\alpha-$direction optical conductivity component from the initial state $m$ and final state $n$.  Explicitly, we have
\begin{align}
&\text{Re}(\sigma_x^{13})  
=\frac{\sigma_0}{8\pi v_z}
\iint dz d\rho 
\Big[4\zeta' (\zeta'z^2+M-\zeta_z' z^2) 
\nonumber\\
&
+\frac{\rho^2}{\rho^2+z^2}[(M-\zeta' \rho^2 -\zeta_z'z^2)^2 
+\varepsilon^2]\Big]
\frac{\rho^3}{\varepsilon^3}
\delta_\Gamma(\omega-2\varepsilon),  
\\
&\text{Re}(\sigma_x^{14})
=\frac{\sigma_0}{4\pi v_z}
\iint dz d\rho 
\frac{\rho z^2}{\varepsilon(\rho^2+z^2)}
\delta_\Gamma(\omega-2\varepsilon), 
\end{align}
and
\begin{align}
&\text{Re}(\sigma_z^{13})=
\frac{\sigma_0v_z}{4\pi v^2}
\iint dz d\rho 
\Big[
4\zeta_z'(M+\zeta_z'\rho^2-\zeta'\rho^2) 
\nonumber\\
&+\frac{1}{\rho^2+z^2}(M-\zeta'\rho^2-\zeta_z' z^2)^2
\Big]\frac{\rho z^2}{\varepsilon^3} \delta_\Gamma(\omega-2\varepsilon), 
\end{align}
\begin{align}
&\text{Re}(\sigma_z^{14})=
\frac{\sigma_0 v_z}{4\pi v^2} 
\iint dz d\rho
\frac{\rho^3}{\varepsilon(\rho^2+z^2)}
\delta_\Gamma(\omega-2\varepsilon),
\end{align}
where $\sigma_0=\frac{e^2}{2\pi}$ is the unit of the quantum conductivity.  As it seems impossible to obtain the analytical results for Re$(\sigma_\alpha)$, we resort to the numerical integration of Eqs.~(6)-(9).  If we use $\frac{\sigma_0}{v_z}$ and $\frac{\sigma_0v_z}{v^2}$ as the unit of Re$(\sigma_x)$ and Re$(\sigma_z)$, respectively, the optical conductivities depend on three quantities $\omega$, $\zeta'$ and $\zeta_z'$. 

\begin{figure*}
	\centering
	\includegraphics[width=18.2cm]{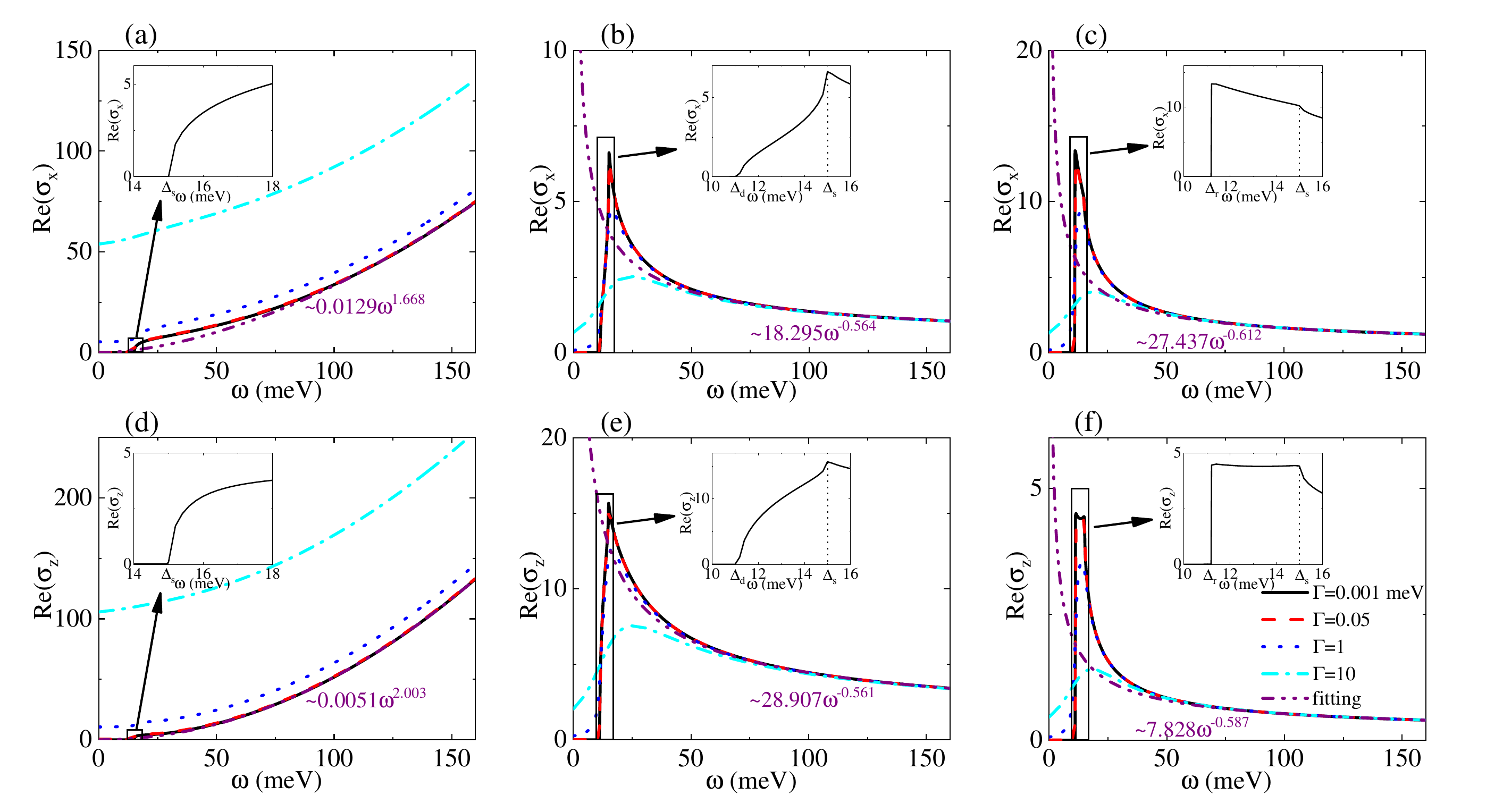}
	\caption{(Color online) The interband optical conductivities Re$(\sigma_x)$ in (a)-(c) and Re$(\sigma_z)$ in (d)-(f) versus the photon frequency $\omega$. Re$(\sigma_x)$ and Re$(\sigma_z)$ are in units of $\frac{\sigma_0}{v_z}$ and $\frac{\sigma_0 v_z}{v^2}$, respectively.  The chosen parameters in different phases are labeled by the asterisks in Fig.~\ref{Fig1} and are the same as Fig.~\ref{Fig2}.  The insets show the enlarged section of the low-energy optical conductivity at the linewidth $\Gamma=0.001$ meV.  We also fit the optical conductivities at high frequency $\omega\gg M$ with the power function Re$(\sigma_x)=a_2\omega^{b_\parallel}$ and Re$(\sigma_z)=a_3\omega^{b_z}$ to extract the exponents $b_\parallel$ and $b_z$.  The different $\Gamma$ are considered in the conductivities and the legends are the same.}
	\label{Fig3}
\end{figure*}

\section{Low-energy features} 

In Figs.~\ref{Fig2} and~\ref{Fig3}, we plot the DOS and optical conductivity, respectively, with the parameters being labeled as the asterisks in Fig.~\ref{Fig1}.  First we consider the limiting clean case, which actually corresponds to the very low linewidth, $\Gamma=0.001$ meV in the numerical calculations.  At low energy $\varepsilon\sim\frac{1}{2}\Delta_s$, the DOS and optical conductivity exhibit distinct features in different phases, which can be seen more clearly in the insets of Figs.~\ref{Fig2} and~\ref{Fig3}, respectively. 

In the single DP phase, the DOS increases smoothly from zero at $\varepsilon=\frac{1}{2}\Delta_s$ to a finite value [Fig.~\ref{Fig2}(a)].  This is also shown in Re$(\sigma_\alpha)$, which increases from zero at $\omega=\Delta_s$ to a finite value [Figs.~\ref{Fig3}(a) and (d)].  
In the double DPs phase, the DOS increases from zero at $\varepsilon=\frac{1}{2}\Delta_d$ to a peak at $\varepsilon=\frac{1}{2}\Delta_s$, and then after a slight drop, it exhibits a power-law like increase with $\varepsilon$ [Fig.~\ref{Fig2}(b)].  Re$(\sigma_\alpha)$ also increases from zero at $\omega=\Delta_d$ to a peak at $\omega=\Delta_s$, but then decreases with $\omega$ [Figs.~\ref{Fig3}(b) and (e)]. 
In the Dirac ring phase, the DOS [Fig.~\ref{Fig2}(c)] and Re$(\sigma_\alpha)$ [Figs.~\ref{Fig3}(c) and (f)] are similar with those in the double DPs phase when $\varepsilon>\frac{1}{2}\Delta_s$ and $\omega>\Delta_s$, respectively.  However, due to the Dirac ring contributions, the DOS begins with a large value at $\varepsilon=\frac{1}{2}\Delta_r$ and Re$(\sigma_\alpha)$ also begins  with a large value at $\omega=\Delta_r$.  The DOS increases linearly till $\varepsilon=\frac{1}{2}\Delta_s$, while Re$(\sigma_x)$ decreases linearly till $\omega=\Delta_s$ and Re$(\sigma_z)$ exhibits a nonlinear change to $\omega=\Delta_s$. 
Thus we suggest that the low-energy DOS and optical conductivities can give clear signatures to distinguish the three different phases. 

When the Fermi energy is away from zero, $\varepsilon_F>\frac{1}{2}\Delta_s$, the low-energy features in Re$(\sigma_\alpha)$ will be screened due to the Pauli blocking effect and the onset of the interband transitions will be $2\varepsilon_F$~\cite{B.Xu,Y.X.Wang2020}.  So, to clearly observe the low-energy features, the low Fermi energy is required and the sample should be lowly doped.

\section{High-energy asymptotic exponents} 

\begin{figure*}
	\centering
	\includegraphics[width=18.2cm]{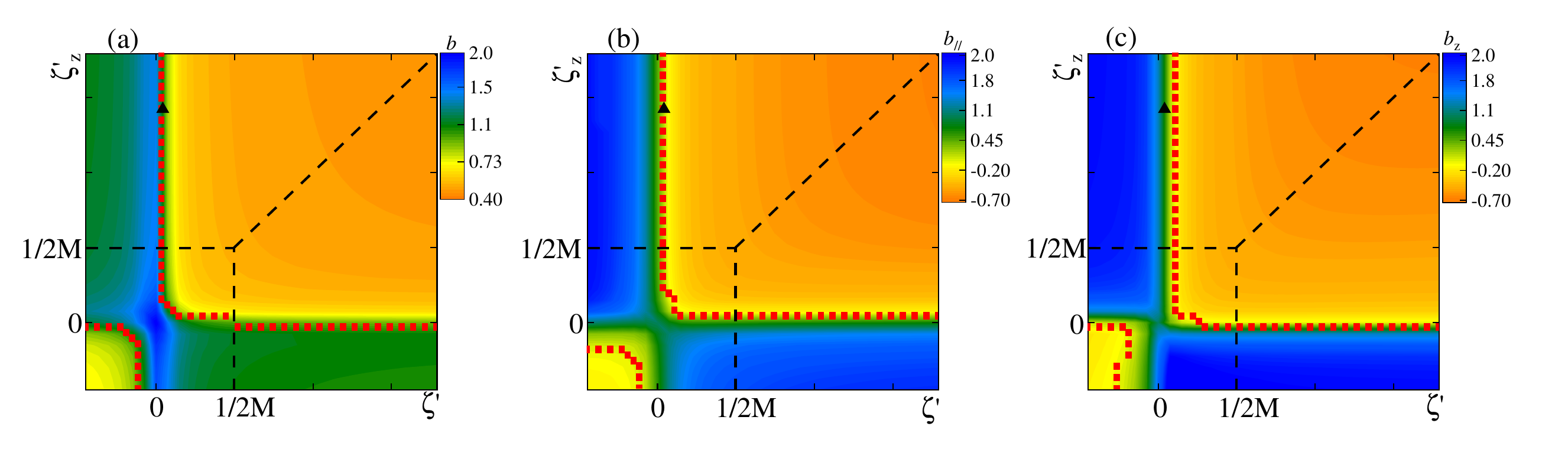}
	\caption{(Color online) The fitted exponents in the parameter space $(\zeta',\zeta_z')$, with $b$ of the DOS in (a), $b_\parallel$ of Re$(\sigma_x)$ in (b) and $b_z$ of Re$(\sigma_z)$ in (c).  The red dotted lines in (a) separate the regions with $b<1$ and $b>1$, in (b) separate the regions with $b_\parallel<0$ and $b_\parallel>0$, and in (c) separate the regions with $b_z<0$ and $b_z>0$.  The dashed lines denote the phase boundaries in Fig.~\ref{Fig1}.  The triangles label the parameters extracted from the experiment~\cite{Y.Jiang} and are close to the red dotted lines. }
	\label{Fig4}
\end{figure*} 

At high energy $\varepsilon\gg M$, the DOS in different phases all exhibit power-law like increase with $\varepsilon$, as shown in Fig.~\ref{Fig2}.  In a linear Dirac semimetal with the dispersion $\varepsilon=\sqrt{v^2(k_x^2+k_y^2+k_z^2)+M^2}$, the DOS is $D(\varepsilon)=\frac{1}{2\pi^2 v^3}\varepsilon\sqrt{\varepsilon^2-M^2}$ and the high-energy characteristic dependence is $D(\varepsilon)\sim\varepsilon^2$.  By using the power function $D(\varepsilon)=a_1\varepsilon^b$, the exponent $b$ is fitted as $b=1.156,0.492,0.472$ in  Figs.~\ref{Fig2}(a)-(c), indicating that the exponent is not a constant, but varies with the model parameters.  The exponents fitted here are smaller than the exponent $2$ in the linear Dirac semimetal.

To further understand the high-energy behavior of the DOS, in Fig.~\ref{Fig4}(a), we present the contour plot of the fitted exponent $b$ in the parameter space $(\zeta',\zeta_z')$.  For $\zeta'=\zeta_z'=0$, the exponent $b=2$, as the system is reduced to the linear Dirac semimetal.  When $\zeta'\zeta_z'>0$, $b$ is around 0.5 and when $\zeta'\zeta_z'<0$, $b$ is around 1.1, meaning that $b$ is largely dependent on the signs of $\zeta'$ and $\zeta_z'$, but not sensitive to the specific phase.  This is easy to understand, as the different phases are characterized by their low-energy features, but not high-energy behaviors.  Note that the regions with $b>1$ and $b<1$ are separated by the red dotted lines.  

Some insights may be gained from the analytic results of the DOS in the clean cases with $\zeta'=0$ and $\zeta_z'=0$~\cite{Supp}.  When $\zeta'=0$, the DOS is 
\begin{align}
&D(\varepsilon)=\frac{\varepsilon}{2\pi^2 v^2v_z}
\Big[z_1\theta(\varepsilon-M)
+(z_1-z_2)\theta(M-\varepsilon)
\nonumber\\
&\times\theta\Big(\varepsilon-\sqrt{\frac{4M\zeta_z'-1}{4\zeta_z'^2}}\Big)
\theta(\zeta_z'-\frac{1}{2M})
\Big], 
\end{align}
where $z_{1/2}=\sqrt{\frac{(2M\zeta_z'-1)\pm\sqrt{4\zeta_z'^2\varepsilon^2-4M\zeta_z'+1}}{2\zeta_z'^2}}$, and $\theta(x)$ is the Heaviside function.  When $\zeta_z'=0$, the DOS is
\begin{align}
&D(\varepsilon)=
\frac{\varepsilon}{8\pi^2 v^2 v_z}
\Big[
\frac{1}{|\zeta'|}
\Big(\frac{\pi}{2}+\text{sgn}(\zeta')\text{arcsin}\frac{M-\frac{1}{2\zeta'}}
{\sqrt{\varepsilon^2+\frac{1-4\zeta'M}{4\zeta'^2} }}\Big)
\nonumber\\
&\times\theta(\varepsilon-M)+\frac{\pi}{\zeta'}
\theta(M-\varepsilon)
\theta\Big(\varepsilon-\sqrt{\frac{4M\zeta'-1}{4\zeta'^2}}\Big)  
\theta(\zeta'-\frac{1}{2M})
\Big]. 
\end{align}
The high-energy DOS is contributed by the term(s) related to $\theta(\varepsilon-M)$ in Eqs.~(10) and (11).  A more careful inspection tells us that the asymptotic exponent $b$ decreases from $\zeta_z'=0$ to large $|\zeta_z'|$ in Eq.~(10) and also decreases from $\zeta'=0$ to large $|\zeta'|$ in Eq.~(11).  Moreover, in Fig.~\ref{Fig4}(a), we have checked that conclusions of Eq.~(10) can be extended to finite $\zeta'$ and the conclusions of Eq.~(11) can be extended to finite $\zeta_z'$. 

On the other hand, for the optical conductivity at high frequency, Fig.~\ref{Fig3} shows that Re$(\sigma_\alpha)$ may increase or decrease with $\omega$.  In the linear Dirac semimetal, Re$(\sigma)$ was revealed to exhibit a linear dependence on $\omega$~\cite{A.Bacsi, P.Hosur, C.J.Tabert} and has been demonstrated in experiments~\cite{R.Y.Chen2015a,B.Xu}. Here we use the power function Re$(\sigma_x)=a_2\omega^{b_\parallel}$ and Re$(\sigma_z)=a_3\omega^{b_z}$ to fit the high-frequency behavior of the optical conductivities in the $x-$ and $z-$direction, respectively.  The fitted exponents are $b_\parallel=1.668,-0.564,-0.612$, and $b_z=2.003,-0.561,-0.587$.  So it is interesting to further investigate the asymptotic behavior of Re$(\sigma_\alpha)$.  In Figs.~\ref{Fig4}(b) and (c), we present the contour plots of $b_\parallel$ and $b_z$ in the parameter space $(\zeta',\zeta_z')$ for Re$(\sigma_x)$ and Re$(\sigma_z)$, respectively.  From the contour plots, several aspects are worthy pointing out: 

(i) When $\zeta'=\zeta_z'$, which means the system is isotropic, the exponents are equal for Re$(\sigma_x)$ and Re$(\sigma_z)$, $b_\parallel=b_z$.  Especially when $\zeta'=\zeta_z'=0$ that the system is reduced to the linear Dirac semimetal, we can see that $b_\parallel=b_z=1$ in Figs.~\ref{Fig4}(b) and (c).  In addition, when $\zeta'=0$ and the maximum $\zeta_z'=0.24$ that the system is close to the 2D conical model, $b_\parallel=0.467$ in Re$(\sigma_x)$ and $b_z=1.281$ in Re$(\sigma_z)$, which are close to the exponents of $0.5$ and $1.5$ in the recent 2D conical model studies~\cite{Y.X.Wang2020, Z.Rukelj}.

(ii) Similar to $b$, the exponents $b_\parallel$ and $b_z$ also depend heavily on the signs of $\zeta'$ and $\zeta_z'$.  The regions with the exponent being positive and negative are separated by the red dotted lines.  We can see that the negative exponents of $b_\parallel<0$ and $b_z<0$ roughly appear in the regions $\zeta'\zeta_z'>0$ [Figs. 4(b) and (c)].  This is because in these regions, both the bands in the $x-y$ plane and $z-$direction are inverted or not inverted simultaneously.  As a result, the DOS owns a smaller exponent $b<1$ [Fig.4(a)] and Re($\sigma_\alpha$) decreases with $\omega$. 

When $\zeta'>0$, in both Re$(\sigma_x)$ and Re$(\sigma_z)$, the exponents decrease with $\zeta_z'$.  But when $\zeta'\leq0$, the behaviors are different: in Re$(\sigma_x)$, the exponent $b_\parallel$ increases with $\zeta_z'$, and in Re$(\sigma_z)$, the exponent $b_z$ increases when $\zeta_z'>0$ but decreases when $\zeta_z'\leq0$. 

\begin{figure}	
	\includegraphics[width=8.2cm]{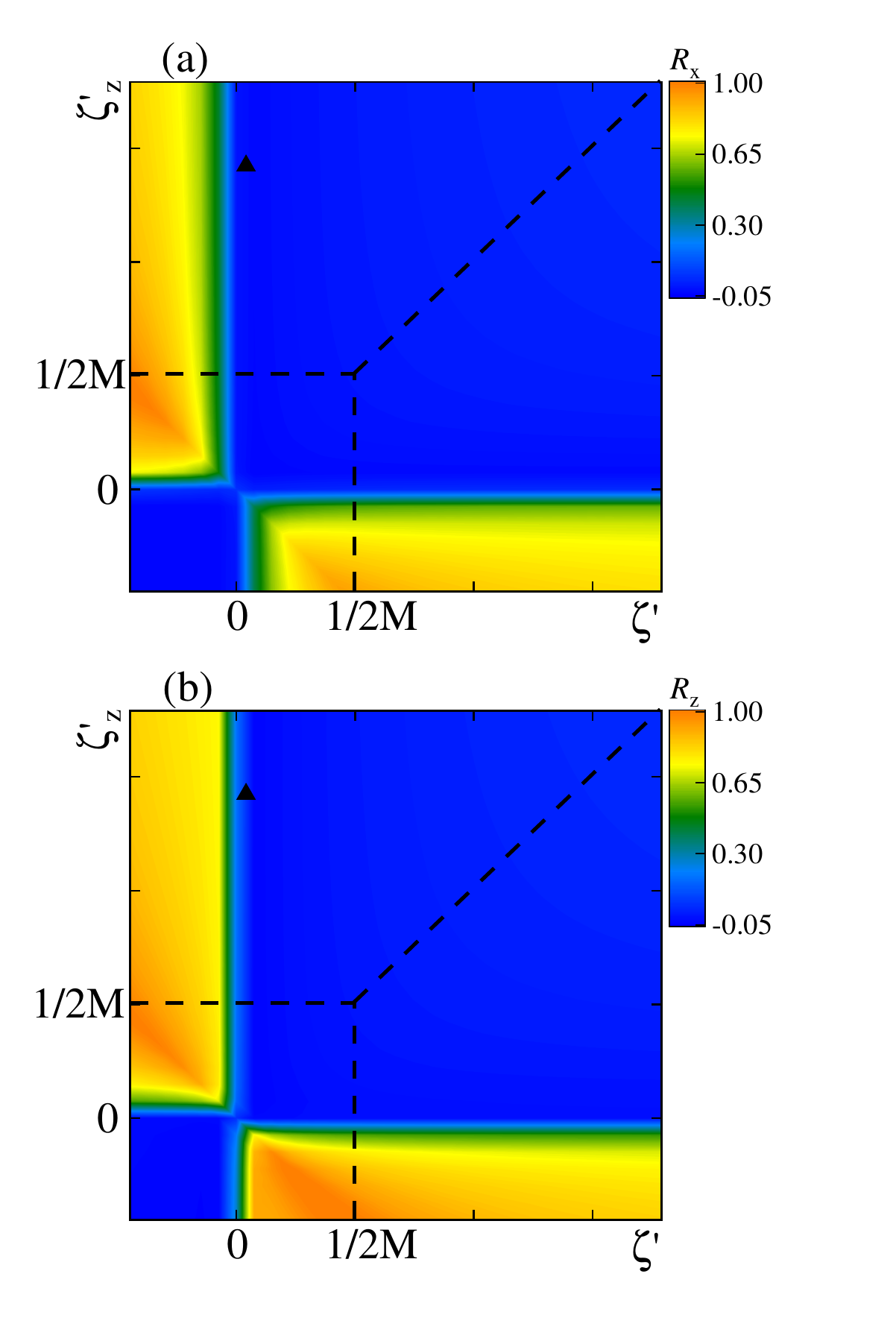}
	\caption{(Color online) The optical conductivity ratio $R_\alpha$ in the parameter space $(\zeta',\zeta_z')$ for $\alpha=x$ in (a) and $\alpha=z$ in (b).  We choose the high photon frequency $\omega=150$ meV.  The black dashed lines denote the phase boundaries.  The triangles label the parameters extracted from the experiment~\cite{Y.Jiang}.}
	\label{Fig5}
\end{figure}

(iii) With the help of the rule-of-thumb formula in the optical conductivity~\cite{M.Dressel}, in which the matrix element of the current density operator is assumed to be a constant, the optical conductivity can be directly related to the DOS $D(\varepsilon)$ as 
\begin{align}
\text{Re}(\sigma_\alpha)\sim \frac{1}{\omega} D(\varepsilon=\frac{\omega}{2}). 
\end{align}
The rule-of-thumb formula holds well for $d-$dimensional Dirac fermions at high energy, as the optical conductivity Re$(\sigma)\sim\omega^{d-2}$~\cite{A.Bacsi} and the DOS $D(\varepsilon)\sim\varepsilon^{d-1}$.  Here based on the numerical results, we demonstrate that for the gapped Dirac semimetal model, the rule-of-thumb formula is satisfied only in the region $(\zeta',\zeta_z')>0$, where both the bands in the $x-y$ plane and $z-$direction are inverted, but fails in other regions.

\section{finite linewidth $\Gamma$} 

To realistically deal with the impurity scatterings, one would apply the self-consistent Born approximation, as it can calculate the impurity-induced self-energy in an effective way ~\cite{C.W.Groth, C.Z.Chen, Y.X.Wang2020b, A.Altland}.  Because the imaginary part of the self-energy will broaden the spectral densities, here we try to phenomenologically capture the effect of the self-energy by considering a finite linewidth $\Gamma$~\cite{P.E.C.Ashby, W.Duan, Y.X.Wang2020, M.Orlita} in the DOS and optical conductivity [Eqs. (4) and (5)].  The results of the DOS and optical conductivity at finite $\Gamma$ are also plotted in Figs.~\ref{Fig2} and ~\ref{Fig3}, respectively.  

At weak $\Gamma=0.05$ meV, both the DOS and Re$(\sigma_\alpha)$ remain unchanged as in the limiting clean case.  With the increasing $\Gamma$, the DOS and Re$(\sigma_\alpha)$ in the gap gradually become nonvanishing and the low-energy features are blurred out, because the bulk electronic states are scattered into the gap by the impurity scatterings.  At $\Gamma=1$ meV, the DOS becomes finite even at the zero energy, indicating that the system has entered the diffusive metal phase~\cite{J.H.Wilson,Y.X.Wang2020b}.  At strong $\Gamma=10$ meV, the DOS at high energy increases a bit in Fig.~\ref{Fig2}(a), but shows certain robustness in Figs.~\ref{Fig2}(b) and (c).  Correspondingly, in Figs.~\ref{Fig3}(a) and (d), Re$(\sigma_\alpha)$ gets evidently enhanced at $\Gamma=1$ meV and will be pushed to a much larger value at strong $\Gamma=10$ meV, while in Figs.~\ref{Fig3}(b), (c), (e) and (f), Re$(\sigma_\alpha)$ at high frequency shows certain robustness to the impurity scatterings, even for the strong $\Gamma=10$ meV. 

To further understand the distinct high-frequency behavior of the optical conductivity at strong $\Gamma=10$ meV, we define the optical conductivity ratio
\begin{align} 
R_\alpha=\frac{\text{Re}[\sigma_\alpha(\Gamma=10\text{ meV})]-\text{Re}[\sigma_\alpha(\Gamma=0.001 \text{ meV})]}
{\text{Re}[\sigma_\alpha(\Gamma=0.001\text{ meV})]}. 
\end{align}
In Fig.~\ref{Fig5}, with the high photon frequency $\omega=150$ meV, the contour plots of $R_\alpha$ are presented in the parameter space $(\zeta',\zeta_z')$ for $\alpha=x$ in (a) and $\alpha=z$ in (b).  We can see that $R_x$ and $R_z$ exhibit similar behaviors and depend heavily on the signs of $\zeta'$ and $\zeta_z'$.  When $\zeta'\zeta_z'>0$, $R_\alpha$ is almost vanishing, meaning that Re$(\sigma_\alpha)$ shows robustness to strong impurity scatterings, which is beneficial for the measurements in experiment.  When $\zeta'\zeta_z'<0$, $R_\alpha$ gives a large value and can reach its maximum $\sim1$ at $\zeta'=-\zeta_z'$, meaning that Re$(\sigma_\alpha)$ is largely enhanced by strong impurity scatterings.

\section{Discussions and Conclusions}

We make some discussions about the implications of our results on experiments and take ZrTe$_5$ as an example.  The band inversion parameters $\zeta'$ and $\zeta_z'$ extracted from the magnetoinfrared spectroscopy measurements in ZrTe$_5$~\cite{Y.Jiang} are labeled as the triangles in Figs.~\ref{Fig4} and~\ref{Fig5}.  For this set of parameters, the asymptotic exponents are obtained as $b=1.437$, $b_\parallel=-0.276$ and $b_z=0.487$ for the DOS, Re$(\sigma_x)$ and Re$(\sigma_z)$, respectively.  Because the triangles lie in different sides of the red dotted lines in Figs.~\ref{Fig4}(b) and (c), Re($\sigma_x$) and Re($\sigma_z$) show opposite asymptotic behaviors, where the former will decrease and the latter will increase with frequency.  These results of the optical conductivity without a magnetic field can provide strong evidences to determine whether the double DPs phase in the proposed gapped Dirac semimetal model is suitable to describe the ground state in ZrTe$_5$.  In addition, as the optical conductivity ratios are as small as $R_x=-0.0154$ and $R_z=0.0719$, the asymptotic optical conductivity can exhibit certain robustness to the impurity scatterings.  Therefore, even though a clean enough sample of ZrTe$_5$ is a requisite for the observation of the low-frequency features, the sample quality is not very strict for observing the high-frequency behaviors.  So more experimental works are expected in the future. 

To summarize, we have made a comprehensive study of the DOS and optical conductivity Re$(\sigma_\alpha)$ in the gapped semimetal model with the mixed linear and parabolic components in all three directions.  We find that the model can support three different phases, which can be distinguished by their low-energy features in the DOS and Re$(\sigma_\alpha)$.  At high energy, the asymptotic exponents for the DOS and Re$(\sigma_\alpha)$, as well as the robustness to the impurity scatterings depend heavily on the signs of $\zeta'$ and $\zeta_z'$.  We believe that more materials whose low-energy excitations can be described by the gapped Dirac semimetal model would be discovered in the future experimental explorations.

\section{Acknowledgments} 

This work was supported by NSFC (Grants No. 11804122 and No. 11905054), and the Fundamental Research Funds for the Central Universities of China.

\end{document}